# Turbulence observations using moored temperature sensors in weakly stratified deep West-Mediterranean waters

## by Hans van Haren[*]

NIOZ Royal Netherlands Institute for Sea Research, P.O. Box 59, 1790 AB  Den Burg, the Netherlands.
[*]Correspnding author: hans.van.haren@nioz.nl
ORCID:0000-0001-8041-8121

**ABSTRACT.** In the stably stratified ocean, small-scale turbulence is important for vertical exchange and hence for the mixing of water masses and suspended matter. To observationally study turbulent motions and the buoyancy- and shear-generators behind them, a 100-m tall array of high-resolution temperature (T) sensors was moored at a 2480-m deep seafloor near the steep continental slope of the Western Mediterranean Sea. The area is dominated by boundary flow, (sub-)mesoscale eddies, internal waves, wintertime dense water formation, and very weak vertical density stratification. Various physical oceanographic processes are observed in detail through a seasonal cycle, including: Low shear-driven turbulence in weak stratification in late-summer and autumn, large buoyancy-driven turbulence by geothermal heating from below alternated with inertial internal wave motions in winter, and moderate buoyancy-like turbulence pushed with stratified waters from above in late-winter and spring.

**INTRODUCTION**

In the vertically density-stratified deep-sea, energy containing turbulent overturning scales have sizes between 1 and 1000 m of duration between 0.01 and 10,000 s. For scientific observational studies on turbulence details such scale-range requires an impossible amount of more than one billion sensors to resolve all processes, as turbulence eventually has a three-dimensional (3D) character in near-homogeneous waters. Because one cannot measure everything, choices have to be made partially based on models.

Thus far, ocean turbulence dissipation rate has been mainly estimated using shipborne 1D-vertical, O(100)-Hz sampling rate microstructure profiling, measuring the smallest details of current-shear variations (e.g., Oakey, 1982; Gregg, 1989), and O(10)-Hz sampling rate CTD-profiling, measuring largest turbulence-energy containing overturns as unstable density portions (Thorpe, 1977). More recently, overturn sizes have been estimated using 1D moored T-sensors, O(10)-m vertical separation at 0.01-Hz sampling rate (Aucan et al., 2006) and O(1)-m vertical interval separation at 1-Hz sampling rate (van Haren and Gostiaux, 2012). Most 1D observations have been made in seas that were vertically well stratified in density, so that: The 100-m large-scale buoyancy frequency amounts N > 10f (f denoting the local inertial frequency), internal waves are abundant, and current-shear dominates vertical (diapycnal) turbulent exchange. 3D quasi-turbulence measurements are difficult to make in the ocean due to logistic problems. Also rare are measurements in deep convection (buoyancy-driven) turbulence areas, like in parts of the Mediterranean Sea in late-winter and spring. In this paper, details are given of several turbulence processes observed by high-resolution moored T-sensors above a deep West-Mediterranean seafloor.

The area is a rare example outside polar seas where deep dense water formation occasionally occurs. This is due to mountain-wind driven winter-drying and -cooling near the surface of relatively warm, salty East-Mediterranean waters transported through the Liguro-Provençal Basin (e.g., Millot, 1999). As a result, apparent density inversions in temperature are expected due to partial salinity-compensation in a complex of various water masses in 1 to 10 km 'sub-mesoscale eddies' (e.g., Testor and Gascard, 2006). During a period of dense-water formation, the continental boundary flow reinforces due to enhanced horizontal density differences (Crépon et al., 1982; Albérola et al., 1995). The stronger flow becomes further unstable resulting in more eddy activity. The effects can reach the deep seafloor.

As tidal motions generally have small amplitudes in the Mediterranean, the dominant internal waves supported by the stratification are found at near-inertial frequencies. In weakly stratified waters internal waves have the property of large vertical columns (Straneo et al., 2002) and may include higher inertial harmonics depending on the value of $N \approx f, 2f, 4f$ (van Haren et al., 2014). Near-inertial waves provide relatively large shear (LeBlond and Mysak, 1978) and can thus contribute to substantial diapycnal turbulent mixing. For the West-Mediterranean, this mixing of heat from above has been compared with convection-turbulence by geothermal heating from below using extensive shipborne microstructure profiling (Ferron et al., 2017). Bethoux and Tailliez (1994) calculated a yearly temperature increase of $6.8 \times 10^{-3}$°C over 100 m above the seafloor due to geothermal heating.

Like convection-turbulence in dense-water formation from the surface, geothermal heating is an effective vertical mixing process. Globally, it contributes about 35 TW of heat flux to the ocean (Davies and Davies, 2010; Wunsch, 2015), which is about 3000 times smaller than incoming solar radiation but about 10 times larger than the kinetic energy in tidal motions. While a common observable in geophysics, geothermal heating is very difficult to observe directly in the deep-sea outside areas of volcanic thermal vents. This is because waters near the seafloor need to be well-conditioned and near-homogeneous. Other processes should not mask convection, such as turbulence-suppressing stratification, internal waves and associated turbulence, and passages of (sub-)mesoscale eddies with varying background temperatures.

In this paper, a seasonal variation will be presented in turbulence observations from moored T-sensors, from internal-wave-dominated low shear-induced turbulence, via moderate convection-turbulence from above, to strong convection-turbulence from below.

**TECHNICAL DETAILS**

Small mooring array of temperature sensors

Netherlands Institute for Sea Research 'NIOZ4' are independent, self-contained high-resolution temperature 'T'-sensors with a precision better than 0.0005°C, a noise level of less than 0.0001°C and a drift of about 0.001°C mo$^{-1}$ after aging (van Haren 2018). Every 4 hours, the internal clocks of all sensors on a mooring, either a 1D single line or a 3D multiple lines mooring, are synchronized via induction to a single standard clock, so that sampling times are less than 0.02 s off.

The compacted five lines (5-L) small 3D-mooring is a 6-m tall, 3-m diameter fold-up, high-grade aluminum structure (van Haren et al., 2016). It consists of two support frames of 1.7x1.7 m each holding a set of four arms 3.3-m long (Figure 1b,c). T-sensors are taped to the five 105-m long, 0.0063-m diameter nylon-coated steel cables. Four instrumented cables connect the corner tips of the upper and lower sets of frame arms 'line-1,…,4'; a central instrumented cable connects the upper and lower inner frames 'line-c'. The corner cables are horizontally separated by 4.0 m from line-c, and by 5.6 (or 8) m from each other. During deployment, the 3D-mooring is stretched after being lifted overboard (Figure 1c). Tension of more than 1 kN per line is achieved first by the central and corner-line weights and, when at the seafloor, by top-floatation providing at least 5 kN total net buoyancy. The 5-L array is dropped to the seafloor in free fall as a common single-line 1D oceanographic mooring.

For the West-Mediterranean deployment, a total available number of 340 NIOZ4 T-sensors is used. 104 T-sensors at 1.0-m intervals are attached to line-c with the lowest sensor at 5 m above the seafloor at z = -2475 m, 53 T-sensors at 2.0 m intervals to line-1,…,4 and 4 T-sensors at 1.0-m interval to each 25-kg corner-line weight so that the lowest sensor is at 0.5 m above the seafloor. For monitoring mooring motions, tilt and compass are activated of 8 T-sensors attached to the tips of the frame-arms. The T-sensors sampled at a rate of 0.5 Hz.

Above the 5-L array, a single line holds acoustic current meter (CM) devices below the main buoyancy elements that keep the mooring as tautly upright as possible. At z = -2310 m a single-point 2-MHz Nortek AquaDopp CM sampled water-flow data once per 150 s.

The mooring was deployed at 42° 47.29′N, 06° 09.11′E, z = -2480 m seafloor, about 40 km south of Toulon-harbor, France, between November 2017 (yearday 321) and September 2018 (yearday 257 + 365 = 622) (Figure 1a). The local bottom slope is nearly flat, <1° from horizontal, but the mooring is only 12 km seaward of the foot of the steep continental slope.

For calibration purposes and to establish the local temperature-density relationship, shipborne CTD-profiles were obtained to within 1 km West from the mooring during the deployment and recovery cruises. The CTD-profile from September 2018 was stopped at -2400 m due to winch constraints. In 2020, a CTD-profile was obtained to within 0.5 m from the seafloor at 6 km northeast of the mooring site.

Moored instruments' performance

The T-sensors generally performed well, but only for the first 4.5 months after deployment. A bad make of batteries caused 50% failure around day 460 and more after that. Less than 10% of the T-sensors collected data during the 10 months underwater. Analysis thus focuses on data from days before 450, when less than 49 (15% of the) T-sensors were either not working, showed calibration problems or were too noisy. Data are interpolated between neighboring sensors. The CM worked fine for the entire 10 months underwater.

Post-processing of moored T-sensor data

For proper instrumental-drift correction not resulting in unrealistic unstable conditions, portions of data are sought in which $d\Theta/dz = 0 < 0.00001°C/100$ m over the entire range of observations as in homogeneous waters. Such homogeneous waters exist in the deep West-Mediterranean as established from previous extensive CTD-observations, especially in the central and southern parts of the basin. These extend up to about 800 m from the seafloor (e.g., van Haren et al., 2014). This extent is two-thirds of the vertical range of 1200 m for geothermal heating dominance suggested by Ferron et al. (2017). Closer to the continental slope, <200-m homogeneous layers are expected due to larger (sub-)mesoscale activity.

In the present records from the 3.5-months of >85% good data, a few periods with a minimum of 30 minutes duration were identified during which environmental temperature variations over 100-m vertical extent were not significantly different than 0.00005°C noise limits. The mean smooth pressure-(adiabatic lapse rate)-corrected zero-slope temperature 'reference' profile for a >30 minute period replaces the mean drift-affected values for each T-sensor in that period. The zero-slope smooth profile is referenced to local CTD-data, for absolute accuracy values. As the drift is time-dependent, corrections for periods away from the zero-slope period are determined using best-fit polynomials of orders that depend on the standard deviation of the drift with respect to the noise level.

**RESULTS**

Monitoring the mooring deployment

The 8 T-sensors at the tips of the frame-arms monitored the deployment, including the overboard-hoisting, the unfolding into full stretching, the sinking to the seafloor, and the final positioning on the seafloor under floatation-tension and mainly horizontal local deep-sea water-flow drag (Figure 2). During down-going free-fall, the initial vertical speed was just above 1 m s$^{-1}$ and reduced to about 0.7 m s$^{-1}$ due to flow-drag during the remainder of the fall (Figure 2a). The environmental temperature (Figure 2b) initially decreased with time, as expected for a statically stable environment, and subsequently increased with time, due to slight pressure (p) compressibility in the weakly stratified waters at $p > 1.5 \times 10^7$ N m$^{-2}$ (Figure 2a). After reaching the seafloor, temperature varied by about $\pm 0.0001°C$ (Figure 2b), which indicates near-homogeneous waters close to instrumental noise levels. The aluminum frame was sturdy and virtually unmoving with the two frames aligned to within 1° (Figure 2c).

CTD-profiles

Between the CTD-profiles considerable variation is seen in Conservative (compressibility corrected; IOC et al., 2010) Temperature $\Theta$ (Figure 3a), with modest changes in density anomaly of less than 0.01 kg m$^{-3}$ (Figure 3b). For z > -1400 m, the stratification is sufficient to support internal waves in a frequency ($\sigma$) band $f \leq \sigma \leq N$ (e.g., LeBlond and Mysak, 1978), which is at least half an order of magnitude wide as N > 4f.

Between -2380 < z < -2180 m, a sign-change is observed in the vertical temperature slope between CTD2017 and -2018. The temperature-density relationship thus shows ambiguous values between the profiles in this vertical range and, most importantly, ambiguous sign.

Between -2480 < z < -2375 m in the near-homogeneous lower 105 m above the seafloor, no sign-change is observed in the temperature-density relationship. Despite differences in stability, the $\Theta$-$\sigma_2$ (density anomaly referenced to $2 \times 10^7$ N m$^{-2}$) consistent relationship amounts over this 105-m vertical range of moored T-sensors,

$$\delta\sigma_2/\delta\Theta = -0.85 \pm 0.05. \tag{1}$$

Over the T-sensors range, N calculated from CTD varies between 0 and 2f (cf. Figure 3b), with a general 100-m scale mean value of about N = 1f = 1.36 cpd (short for cycle per day) and maxima of small-1-m-scale buoyancy frequency $N_s$ up to about $N_{s,max}$ = 4f. The vertical range of N = 0 (homogenous, no stratification) is <100 m from the seafloor in these profiles.

As a result, the general mean range for the inertio-gravity wave (IGW) band is [$\sigma_{min}$, $\sigma_{max}$] = [0.6f, 1.7N] = [0.8, 2.3] cpd. IGW-limits are computed as [$\sigma_{min}$, $\sigma_{max}$] = $1/\sqrt{2} \cdot [(A-B)^{1/2}, (A+B)^{1/2}]$ where A = $N^2 + f^2 + f_s^2$ and B = $(A^2 - (2fN)^2)^{1/2}$, which results in limits below f and above N, respectively (e.g., LeBlond and Mysak, 1978). Knowledge of the weak stratification is thus important for studies on particular internal wave dynamics, which can only be achieved via careful post-processing applying corrections to the moored T-sensor data.

The relatively tight and consistent relationship (1) implies the T-sensor data may be used as a proxy for density variations to infer turbulence values using the method of reordering unstable data-points to monotonously stable vertical profiles (Thorpe, 1977).

These overturns follow after reordering every 2 s the 105-m high potential density (Conservative Temperature) profile $\sigma_2(z)$, which may contain inversions, into a stable monotonic profile $\sigma_2(z_s)$ without inversions. After comparing observed and reordered profiles, displacements d = min(|z-$z_s$|)·sgn(z-$z_s$) are calculated necessary for generating the reordered stable profile. Certain tests apply to disregard apparent displacements associated with instrumental noise and post-calibration errors (Galbraith and Kelley, 1996). Such a test-threshold is very low for NIOZ T-sensor data, <0.0005°C. Then,

$$\varepsilon = 0.64d^2N^3, \tag{2}$$

the turbulence dissipation rate, where N is computed from each of the reordered, essentially statically stable, vertical density profiles. The numerical constant follows from empirically relating the rms overturning scale $d_{rms}$ with the Ozmidov-scale ($L_O$) of largest isotropic turbulence overturns in a stratified fluid: $L_O/d_{rms}$ = 0.8 (Dillon, 1982).

This ratio reflects turbulence in any high Reynolds number stably stratified environment like the deep ocean, in which shear-driven and convection turbulence intermingle and difficult to separate. Comparison between calculated turbulence values using shear measurements and using Thorpe overturning scales with $L_O$=0.8$d_{rms}$ led to 'consistent results' (Nash et al., 2007).

In (2), individual d are used rather than taking their rms-value across a single overturn as originally proposed by Thorpe (1977). The reason is that individual overturns cannot easily be distinguished, first, because they are found at various scales with small ones overprinting larger overturns, and, second, because some overturns exceed the range of T-sensors. Therefore, 'mean' turbulence values are calculated by arithmetic averaging $\varepsilon(t, z)$ from (2) in the vertical […] or in time <…>, or both. Although Θ-data are analyzed throughout, 'temperature' is used in their description for short henceforth.

Environmental variability with time and frequency

The mooring remained almost unaffected by environmental conditions, also when water-flow speeds went up to 0.35 m s$^{-1}$ in late winter (Figure 4a): The mooring rotated never more than ±3°. In temperature, a single 10-month T-record varied by <0.01°C, with most variation in a limited number of positive peaks between days 430 (early March) and 560 (mid-July). One-third of the temperature variation was attributable to instrumental drift causing a low-frequency nonlinear increase with time, mainly during the first month (Figure 4b).

In vertical temperature difference ΔΘ, the positive peaks reflecting stable stratification stand out, not only over the 100-m vertical scale but also to smaller extent over 16-m vertical scale near the seafloor (Figure 4c). Both 10-month records also show significant unstable values. Largest negative temperature differences are observed in early winter, well before the occurrence of the positive peaks. For given fixed relationship (1), the negative ΔΘ-values indicate rather persistent large-scale instabilities, possibly related to convection. For ΔΘ = -0.001°C, e.g. between days 350 and 400 in Figure 4c, this translates to a vertical range of weak instability of 250 m from the seafloor in CTD-temperature albeit only of about 150 m in density anomaly, both well exceeding the vertical range of moored T-sensors.

The spectral overview (Figure 4d) for the first 3.5 months of T-sensor data demonstrates the kinetic energy of local horizontal water-flows is dominated by a peak at f (about 17.5-h periodicity) and larger energy at sub-f (about 10-day periodicity). The vertical water-flow component shows a small but significant peak-bulge in variance around f, confirming rotation-driven IGW in very weakly stratified waters, and semidiurnal tide.

The mean T-variance spectra do not show significant spectral peaks, but several spectral slopes that vary with frequency (band) and vertical position. Within the IGW-band, spectral slopes for both upper and lower sensor data scale with σ$^{-1}$, as observed in the open ocean (van

Haren and Gostiaux, 2009). For the upper sensors in $\sigma_{max} < \sigma < 300$ cpd, the slope scales approximately with $\sigma^{-5/3}$, reflecting the inertial subrange of (shear-induced) turbulence for a passive scalar (Tennekes and Lumley, 1972; Warhaft, 2000), before rolling off to noise levels. Although this range includes the small-2-m-scale buoyancy frequency $N_s$, the range $N < N_s < N_{s,max}$ describing internal waves in thin stratified layers, the slope does not reflect freely propagating internal waves that scale with $\sigma^{-2}$ (Garrett and Munk, 1972).

For the lower sensors, the T-variance demonstrates about one order of magnitude smaller values compared with the upper sensors. For $\sigma_{max} < \sigma < N_{s,max}$, the slope scales with $\sigma^{-2.5}$, representing either internal waves or, more likely, finestructure contamination (Phillips, 1971). For $N_{s,max} < \sigma < 120$ cpd, the slope scales with $\sigma^{-7/5}$, which reflects an active scalar (Bolgiano 1959; Pawar and Arakeri 2016) and convection-turbulence. Thus spectrally, T-sensors show a limited IGW, and extended finestructure and inertial subrange of turbulence away from the seafloor, with convection turbulence near the seafloor.

In terms of small-scale comparison between all independent pairs of T-sensors for the 103-day period between days 322 and 425 in winter, strong coherence is found up to $N_{s,max}$, before rolling off to insignificant values (Figure 4e). The 8-m horizontal scale matches that of the 4-m (and even smaller) vertical scale, suggesting vertically restricted motions. For ~150 < $\sigma < 300$ cpd, the 8-m horizontal scale coherence matches that of the 8-m vertical scale coherence, which suggests isotropic motions. In the range in between, $N_{s,max} < \sigma < 150$ cpd, the convection-turbulence observed in the lower T-sensors marks the transition from anisotropic 'stratified' to isotropic turbulence.

**DYNAMICS DETAILS**

Moored T-sensor data highlight several dynamical features of the seasonal variation in stability near the seafloor close to the continental slope of the deep West-Mediterranean.

Mainly stable stratification, deep internal waves, and some turbulence

In summer and autumn, between about days 210 and 340, the waters near the T-sensor mooring site are weakly but stably stratified and double-inertial period internal waves can appear close to the seafloor (Figure 5a). Turbulence activity is as low as in open-ocean interior waters (Figure 5b), with 1-d, 105-m mean $[<\varepsilon>] = 2\pm1\times10^{-10}$ m$^2$s$^{-3}$ for $[<N>] = 2.0\pm0.2\times10^{-4}$ s$^{-1} \approx 2f$. Nevertheless, waters are not quiescent as in laminar flows and bursts of turbulence occasionally appear (e.g., on day 340.3). No convection-turbulence is observed from the seafloor reaching upward. The stable vertical density stratification apparently masks any direct observation of geothermal heating.

Mainly apparently stable stratification and considerable turbulent bursts

The stable $N=2f$ stratification is built-up in the preceding spring, between about late-winter-day 55 (430 in the following year) and mid-summer day 210 (575). During this period, deep water-flows reach speeds >0.1 m s$^{-1}$, up to 0.35 m s$^{-1}$ (Figure 4a), and which are occasionally, but not always, accompanied by positive peaks in temperature (Figure 4b). The peaks reflect warm waters coming from above, possibly direct via convection, or being pushed from above, either via convection or internal waves against stable stratification (Figure 6). The period coincides with the common period of dense water formation near the surface, which reaches several 100's of meters deep every year but the seafloor once every 2-5 years, approximately (Mertens and Schott, 1998; Somot et al., 2018). The local stable stratification peaks to $N \approx 4f$, but quasi-convection turbulence is obvious. This convection has been attributed to internal wave push and active turbulence activity that overcome the local reduced gravity in the lower 50 m above the floor of Lake Garda where mean $N = 2f$ (van Haren and Dijkstra, 2021). In that 340-m deep alpine lake, convection-turbulence from above is observed below strongly stratified waters. For the present data one cannot rule out the partial density-compensation effects of salinity in different water masses (from above), which may cause ambiguous temperature-density relationship due to dense water formation in late-winter and spring. Adopting a conservative stand, no turbulence values are presented for this period. (If turbulence values are calculated using (2), they fall between those of Sections above and below). Such short warm-water blob coming from above is surrounded by unstable waters over 105 m above the seafloor, see the leftmost part of Figure 6. Note that the T-range is about 0.0045°C and that convection-detection generally requires smaller ranges.

Mainly apparently unstable stratification turbulence

In winter between days 340 and 430, waters near the seafloor are sufficiently weakly stratified so that presumed geothermal heating may be directly observable (Figure 7). Also during this period, deep dense water formation is not yet active as near-surface waters are not sufficiently cooled and dried to become denser than all underlying waters. Thus, convection of warmer waters from above that may mask geothermal convection from below are not expected before the end of winter.

The 0.0011°C-range in Figure 7a demonstrates warmest waters near the seafloor for half an inertial period, with typically 'plumes' of alternating (upgoing) warm and (downgoing) cooler waters depicting natural convection. The individual convection plumes are variable in temporal and vertical extent, and in intensity. They resemble the irregular plumes of opposite sign in Figure 6. In Figure 7, the heating of waters above varies with time, as temperature is not constant near the seafloor. The convection from below is surrounded by stably stratified waters about half an inertial period (0.38-d) apart. Using conventional Thorpe-scale analysis (2), the 0.54-d, 105-m mean $[<\varepsilon>] = 2.2\pm1.5\times10^{-8}$ m$^2$s$^{-3}$ for $[<N>] = 2.0\pm0.2\times10^{-4}$ s$^{-1} \approx 2f$.

Recall that the mean 105-m-scale buoyancy frequency is determined from the reordered, stably stratified vertical profiles. Its absolute value is the same to within error as the mean values for the stratified periods. This suggests that convection-turbulence is working on the same stratification as shear-induced turbulence might have.

Averaging data from Figure 7 only over the period of convection, the 0.38-d, 105-m mean $[<\varepsilon>] = 2.8\pm1.6\times10^{-8}$ m$^2$s$^{-3}$ for above $[<N>]$. The variation with time is considerable in up- and down-going motions, but only about one order of magnitude in vertically averaged dissipation rate. Integrated over the 105-m range of observations however, the turbulence energy dissipation rate amounts 3 mW m$^{-2}$ after conversion with density of seawater. This value is just 3% of the average amount of the vertical energy flux attributed to geothermal heating in the area, which is 100±30 mW m$^{-2}$ (Pasquale et al., 1996). Even if one considers a turbulent mixing efficiency of about 0.5, which is typical for vertical natural convection (Ng et al., 2016), Rayleigh-Taylor instabilities in high Reynolds number flows (Dalziel et al., 2008) and Rayleigh-Bénard convection (Gayen at al., 2013), the here calculated turbulence dissipation rates fall short by one order of magnitude of a presumed value of 50 mW. Two suggestions are given for the apparent discrepancy:

1. The moored array has not resolved the vertical extent of the convection. Judging from CTD-data, the vertical density gradient becomes steep enough so that large-scale N > 2f at z = -2225±75 m. Thus, assuming convection reaches with the same turbulence intensity that far from the seafloor, the integration over 105 m could be extended by 150±75 m. However, this only yields turbulence dissipation being 7.5% of the average geothermal heat flux. To reach 50% one requires a vertical integration over about 1700 m from the seafloor, which would be reaching well within the strongly stratified waters near the surface (Figure 3b).

2. One may argue that convection has to overcome the stratification induced by the strongest small-scale stratification, e.g. found in thin layers along sides of plumes (Li and Li 2006). Per vertical profile one can determine the maximum $N_{s,max}(t)$ of $N_s$ and use (2) to compute, per time step,

$$\varepsilon_s = 0.64d^2 N_{s,max}{}^3, \qquad (3)$$

which gives after averaging over the 0.38-d period and integrating over the 105-m vertical range: 40 mW m$^{-2}$. This value is within the range of error of geothermal heat flux assuming a mixing efficiency of 0.5.

A combination of points 1 and 2 may provide the required 50 mW for a vertical integration range of 130 m from the seafloor.

**DISCUSSION AND CONCLUSIONS**

The high-resolution temperature observations from moored T-sensors demonstrate active turbulence and internal wave processes in very weakly stratified waters of the deep Western Mediterranean. The T-sensor range of 0.5-105 m above the flat seafloor in the vicinity of the steep continental slope demonstrates mean stratification so that $0 < N < 2f$, with $N > 2f$ higher up. Bethoux and Tailliez (1994) calculated a homogenization of a 400-m thick layer with initial temperature gradient of 0.00001°C m$^{-1}$ due to geothermal heating over one year. For the present site (within one year), initial $N = 3f$ would reduce to $N = 2f$ in a 160-m thick layer, as in CTD-data. Over the T-sensor range, a consistent temperature-density relationship is found, also under unstable conditions. As a result, temperature can be used as tracer for density and turbulence values can be calculated. A seasonal cycle in temperature instability and thus in turbulence values is suggested from the observations near the deep seafloor.

In winter, instabilities and convection turbulence from below are largest, when water-flow amplitudes are smaller than 0.1 m s$^{-1}$, dominated by near-inertial motions. The convection-turbulence from below observed by T-sensors is mainly attributed to geothermal heating. Although extending above the 105-m range of moored T-sensors, CTD-observations suggest it is limited to the lower 200 m above the seafloor, rather than 1200 m as suggested for the open West-Mediterranean basin (Ferron et al., 2017). The overall geothermal warming of the deep waters by about +0.006°C over 10 months (Bethoux and Tailliez, 1994) cannot be confirmed from the present moored T-sensor data, because of unknown instrumental drift and because of unknown cooling from above or continental shelf cascading (CTD2017 and -2018 showed a difference of -0.005°C, also with unknown instrumental drift).

Throughout spring, geothermal heating from below is alternated with apparently stable warmer waters that are pushed down, possibly driven by convection processes near the surface and/or (sub)mesoscale eddies, also occurring with inertial and semi-inertial periodicities. Deep water-flow speeds regularly double in size compared to winter. Convection from above seems turbulent also compared to geothermal convection from below.

By early summer, geothermal instabilities become smaller in time and weaker in amplitude, presumably dampened by the added heat (stratification) from above.

In late-summer and autumn, convection activities are low, both from below and above. The stratification of typically N = 2f reaches close to the seafloor, dampens the turbulent exchange and supports internal waves of considerable O(10)-m amplitudes.

Throughout autumn, stratification becomes gradually eroded, presumably by geothermal heating that remains undetectable in the stratified waters.

**ACKNOWLEDGMENTS**

This research was supported in part by Netherlands Organization for Scientific Research (NWO). I acknowledge captain and crew of the R/V l'Atalante and NIOZ-MRF for their very helpful assistance during deployment and recovery of the mooring. I thank V. Bertin, CPPM-Marseille and ANTARES, for logical assistance and M. Stastna (Univ. Waterloo, Canada) for providing the 'darkjet' colour-map suited for T-sensor data.

**REFERENCES**

Albérola, C., C. Millot, C., and J. Font. 1995. On the seasonal and mesoscale variabilities of the Northern Current during the PRIMO-0 experiment in the western Mediterranean Sea. *Oceanologica Acta* 18, 163-192, https://explore.openaire.eu/search/publication?articleId=od_________7::ce5c7bcab260aa735747dce7adeb709b.

Aucan, J., M.A. Merrifield, D.S. Luther, and P. Flament. 2006. Tidal mixing events on the deep flanks of Kaena Ridge, Hawaii. *Journal of Physical Oceanography* 36, 1202-1219, https://doi.org/10.1175/JPO2888.1.

Bethoux, J.P., and D. Tailliez. 1994. Deep-Water in the Western Mediterranean Sea, Yearly Climatic Signature and Enigmatic Spreading. In: *Ocean Processes in Climate Dynamics: Global and Mediterranean Examples*, Malanotte-Rizzoli, P., and A.R. Robinson (Eds), Kluwer Academic Publishers, 355-369.

Bolgiano, R. 1959. Turbulent spectra in a stably stratified atmosphere. *Journal of Geophysical Research* 64, 2226-2229, https://doi.org/10.1029/JZ064i012p02226.

Crépon, M., L. Wald, and J.M. Monget. 1982. Low-frequency waves in the Ligurian Sea during December 1977. *Journal of Geophysical Research* 87, 595-600, https://doi.org/10.1029/JC087iC01p00595.

Dalziel, S.B., M.D. Patterson, C.P. Caulfield, and I.A. Coomaraswamy. 2008. Mixing efficiency in high-aspect-ratio Rayleigh-Taylor experiments. *Physics of Fluids* 20, 065106, https://doi.org/10.1063/1.2936311.

Davies, J.H., and D.R. Davies. 2010. Earth's surface heat flux. *Solid Earth* 1, 5-24, https://se.copernicus.org/articles/1/5/2010/se-1-5-2010.pdf.

Dillon, T.M. 1982. Vertical overturns: a comparison of Thorpe and Ozmidov length scales. *Journal of Geophysical Research* 87, 9601-9613, https://doi.org/10.1029/JC087iC12p09601.

Ferron, B., P. Bouruet Aubertot, Y. Cuypers, K. Schroeder, and M. Borghini. 2017. How important are diapycnal mixing and geothermal heating for the deep circulation of the

Western Mediterranean? *Geophysical Research Letters* 44, 7845-7854. doi:10.1002/2017GL074169.

Galbraith, P.S., and D.E. Kelley. 1996. Identifying overturns in CTD profiles. *Journal of Atmospheric and Oceanic Technology* 13, 688-702, https://doi.org/10.1175/1520-0426(1996)013<0688:IOICP>2.0.CO;2.

Garrett, C., and W. Munk. 1972. Space-time scales of internal waves. *Geophysical Fluid Dynamics* 3, 225-264, https://doi.org/10.1080/03091927208236082.

Gayen, B., G.O. Hughes, and R.W. Griffiths. 2013. Completing the mechanical energy pathways in turbulent Rayleigh-Bénard convection. *Physical Review Letters* 111, 124301, https://link.aps.org/doi/10.1103/PhysRevLett.111.124301.

Gregg, M.C. 1989. Scaling turbulent dissipation in the thermocline. *Journal of Geophysical Research* 94, 9686-9698, https://doi.org/10.1029/JC094iC07p09686.

IOC, SCOR, IAPSO, 2010: *The international thermodynamic equation of seawater – 2010: Calculation and use of thermodynamic properties*. Intergovernmental Oceanographic Commission, Manuals and Guides No. 56, UNESCO, 196 pp.

LeBlond, P.H., and L.A. Mysak. 1978. *Waves in the Ocean*. Elsevier, 602 pp.

Li, S., and H. Li. 2006. Parallel AMR code for compressible MHD and HD equations. T-7, MS B284, Theoretical division, Los Alamos National Laboratory, http://citeseerx.ist.psu.edu/viewdoc/summary;jsessionid=1548A302FD5C2B1DFAC1BA7A5E70605F?doi=10.1.1.694.3243, last accessed 21 September 2022.

Mertens, C., and F. Schott. 1998. Interannual variability of deep convection in the Northwestern Mediterranean. *Journal of Physical Oceanography* 28, 1410-1424, https://doi.org/10.1175/1520-0485(1998)028<1410:IVODWF>2.0.CO;2.

Millot, C., 1999. Circulation in the Western Mediterranean Sea. *Journal of Marine Systems* 20, 423-442, https://doi.org/10.1016/S0924-7963(98)00078-5.

Nash, J.D., M.H. Alford, E. Kunze, K. Martini, and S. Kelly. 2007. Hotspots of deep ocean mixing on the Oregon. *Geophysical Research Letters* 34, L01605, doi:10.1029/2006GL028170.

Ng, C.S., A. Ooi, and D. Chung. 2016. Potential energy in vertical natural convection. Proc. 20[th] Australasian Fluid Mech. Conf., 727 (4 pp), https://people.eng.unimelb.edu.au/imarusic/proceedings/20/727%20Paper.pdf.

Oakey, N.S. 1982. Determination of the rate of dissipation of turbulent energy from simultaneous temperature and velocity shear microstructure measurements. *Journal of Physical Oceanography* 12, 256-271, doi:10.1175/1520-0485(1982)012<0256:dotrod>2.0.co;2.

Pasquale, V., M. Verdoya, and P. Chiozzi. 1996. Heat flux and timing of the drifting stage in the Ligurian–Provençal basin (NWMediterranean). *Journal of Geodynamics* 21, 205-222, https://doi.org/10.1016/0264-3707(95)00035-6.

Pawar, S.S., and J.H. Arakeri. 2016. Kinetic energy and scalar spectra in high Rayleigh number axially homogeneous buoyancy driven turbulence. *Physics of Fluids* 28, 065103, https://doi.org/10.1063/1.4953858.

Phillips, O.M. 1971. On spectra measured in an undulating layered medium. *Journal of Physical Oceanography* 1, 1-6, https://doi.org/10.1175/1520-0485(1971)001<0001:OSMIAU>2.0.CO;2.

Somot, S., et al. 2018. Characterizing, modelling and understanding the climate variability of the deep water formation in the North-Western Mediterranean Sea, Climate Dynamics 51, 1179-2010, doi:10.1007/s00382-016-3295-0.

Straneo, F., M. Kawase, and S.C. Riser. 2002. Idealized models of slantwise convection in a baroclinic flow. *Journal of Physical Oceanography* 32, 558-572, https://doi.org/10.1175/1520-0485(2002)032<0558:IMOSCI>2.0.CO;2.

Tennekes, H., and J.L. Lumley. 1972. *A first course in turbulence*. MIT Press, 320 pp.

Testor, P., and J.-C. Gascard. 2006. Post-convection spreading phase in the Northwestern Mediterranean Sea. *Deep-Sea Research I* 53, 869-893, https://doi.org/10.1016/j.dsr.2006.02.004.

Thorpe, S.A. 1977. Turbulence and mixing in a Scottish loch. *Philosophical Transactions of the Royal Society of London A* 286, 125-181, https://doi.org/10.1098/rsta.1977.0112.

van Haren, H. 2018. Philosophy and application of high-resolution temperature sensors for stratified waters. *Sensors* 18, 3184, doi:10.3390/s18103184.

van Haren, H., and H.A. Dijkstra. 2021. Convection under internal waves in an alpine lake. *Environmental Fluid Mechanics* 21, 305-316, https://doi.org/10.1007/s10652-020-09774-2.

van Haren, H., and L. Gostiaux. 2009. High-resolution open-ocean temperature spectra. *Journal of Geophysical Research* 114, C05005, doi:10.1029/2008JC004967.

van Haren, H., and L. Gostiaux. 2012. Detailed internal wave mixing observed above a deep-ocean slope. *Journal of Marine Research* 70, 173-197, https://doi.org/10.1357/002224012800502363.

van Haren, H. et al. 2014. High-frequency internal wave motions at the ANTARES site in the deep Western Mediterranean. *Ocean Dynamics* 64, 507-517, https://doi.org/10.1007/s10236-014-0702-0.

van Haren, H., J. van Heerwaarden, R. Bakker, and M. Laan. 2016. Construction of a 3D mooring array of temperature sensors. *Journal of Atmospheric and Oceanic Technology* 33, 2247-2257, https://doi.org/10.1175/JTECH-D-16-0078.1.

Warhaft, Z. 2000. Passive scalars in turbulent flows. *Annual Review of Fluid Mechanics* 32, 203-240, https://doi.org/10.1146/annurev.fluid.32.1.203.

Wunsch, C. 2015. *Modern observational physical oceanography—Understanding the global ocean*. Princeton University Press, 493 pp.

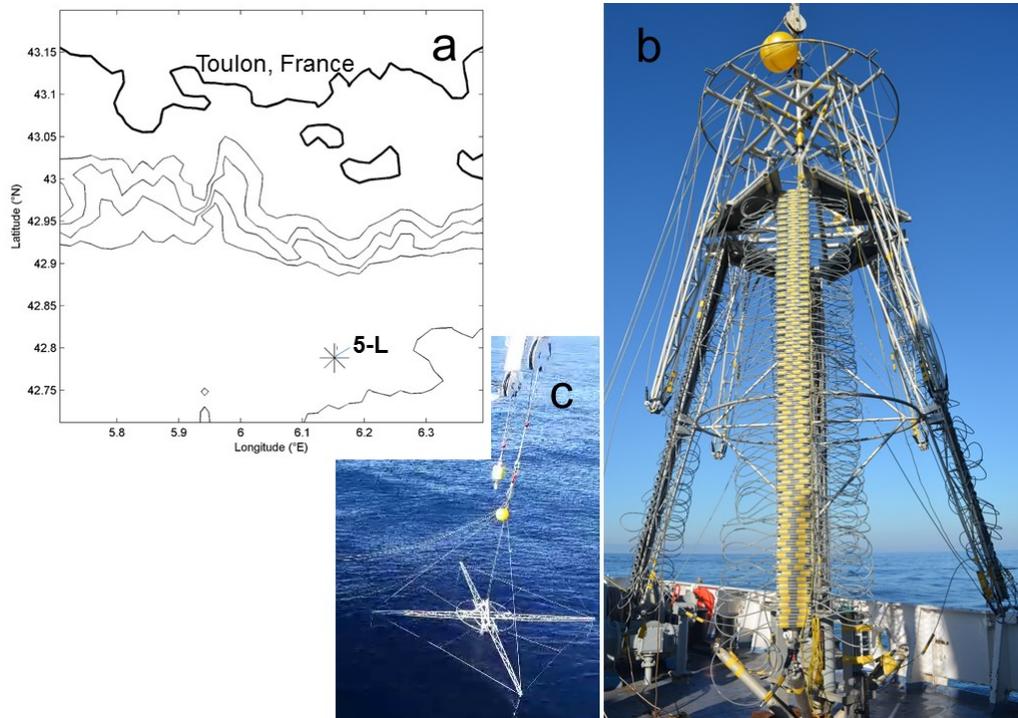

**Figure 1**. Five lines (5-L) mooring frame and deployment site. (a) Location of the 5-L (star) about 40 km south of Toulon, France, and about 12 km south of the continental slope. Depth-contours are drawn every 500 m. (b). 5-L folded-up and ready for deployment on board of R/V l'Atalante. (c). 5-L unfolded overboard, just before release in free-fall.

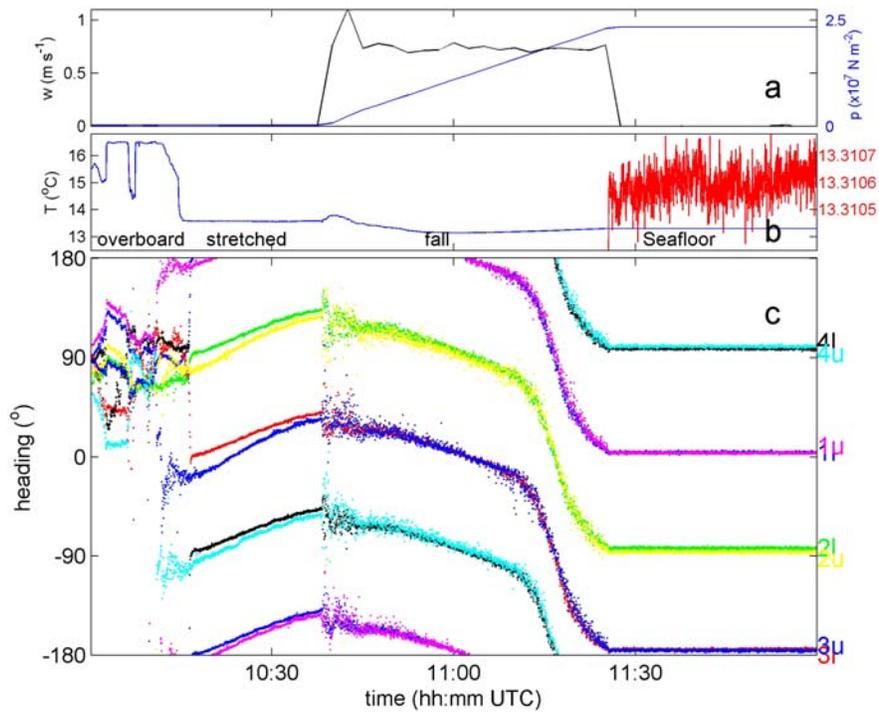

**Figure 2**. Operational observations during deployment of 5-L-mooring: From overboard hoisting, stretching, descend in free-fall, to stationary position at the seafloor. (a) Vertical velocity (black; scale to the left; positive is when the mooring descends) and pressure (blue; scale to right) registered every 150 s by the current meter (CM) 170 m above the central weight. (b) Temperature (blue; scale to left) and its 10,000-times magnification arbitrarily shifted vertically (red; scale to right). (c) Compass-heading of all eight mooring-frame corner-sensors of upper (u) and lower (l) aluminum frames.

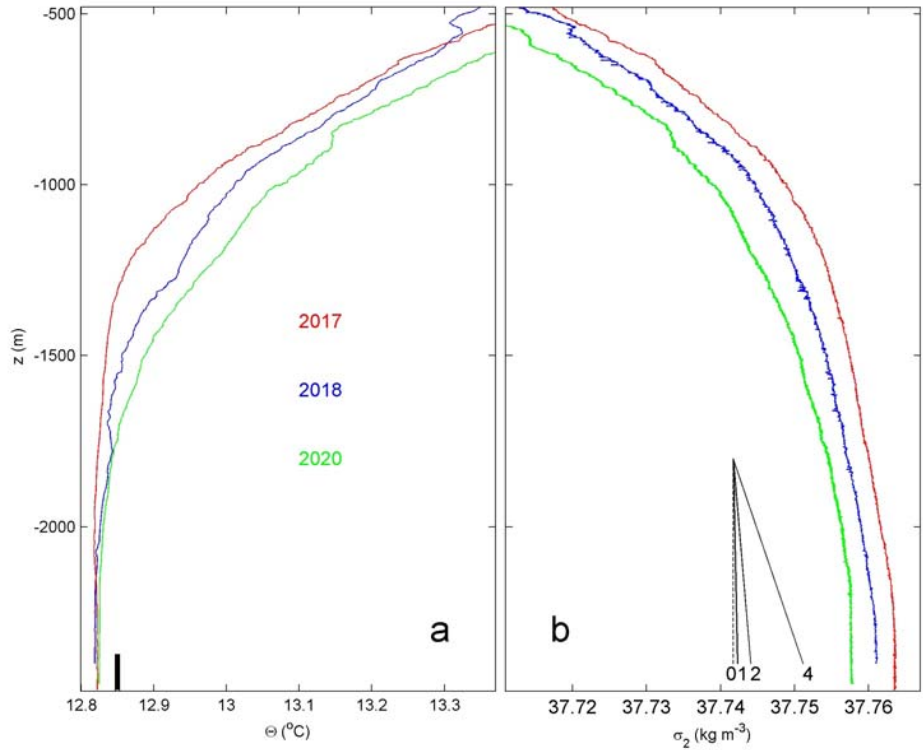

**Figure 3**. Lower 2000 m of shipborne CTD-profiles obtained near the 5-L-mooring site during deployment (red; with lowest value 5 m above the local seafloor) and recovery (blue; with lowest value 80 m above the seafloor) cruises with an extra profile from October 2020 (green; with lowest value 0.5 m above the seafloor). (a) Conservative Temperature. The vertical bar indicates the extent of the moored T-sensors. (c) Density anomaly referenced to $2 \times 10^7$ N m$^{-2}$. Four vertical slopes are shown equivalent to buoyancy (N)/local inertial (f) frequency ratio N/f = 0, 1, 2, 4.

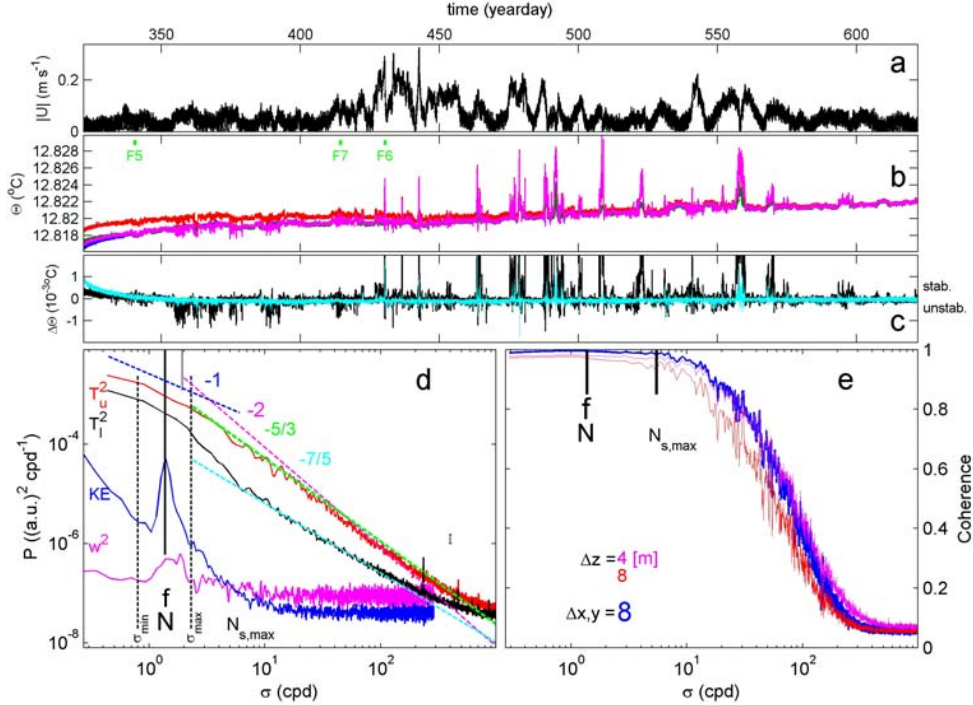

**Figure 4**. Overview of 10 months of mooring data. (a) Horizontal water-flow amplitude observed by CM at z = -2310 m. (b) Calibrated but not drift-corrected Conservative Temperature from -2475 (blue), -2459 (green; barely visible), -2393 (red) and -2377 m (purple) of line-1. Green ticks indicate days of F(igures) 5-7. (c) Temperature difference between records from b.: from -2459 minus -2475 m (light-blue), and from -2377 minus -2475 m (black). Data are low-pass filtered with cut-off at 1000 cpd. (d) Moderately smoothed spectra for temperature variance from the period between days 322 and 425 in b., averages for lower two T-sensors (black) and upper two (red). For reference, CM-data provide arbitrarily vertically shifted horizontal kinetic energy 'KE' and vertical current variance '$w^2$'. The small vertical line indicates the semidiurnal lunar tidal frequency and the long vertical dashed lines indicate the non-traditional inertio-gravity wave (IGW) band for weakly stratified waters with N = f. $N_{s,max}$ = 4f denotes the average of maximum small-2-m-scale buoyancy frequencies per profile. Several spectral slopes are given in the log-log plot with, e.g., '-1' indicating $\sigma^{-1}$ (see text). (e) Coherence spectra between all possible pairs of T-sensors from 5-L across indicated vertical and horizontal distances, average for days 322 to 425. The 95% significance level is at about coherence = 0.05.

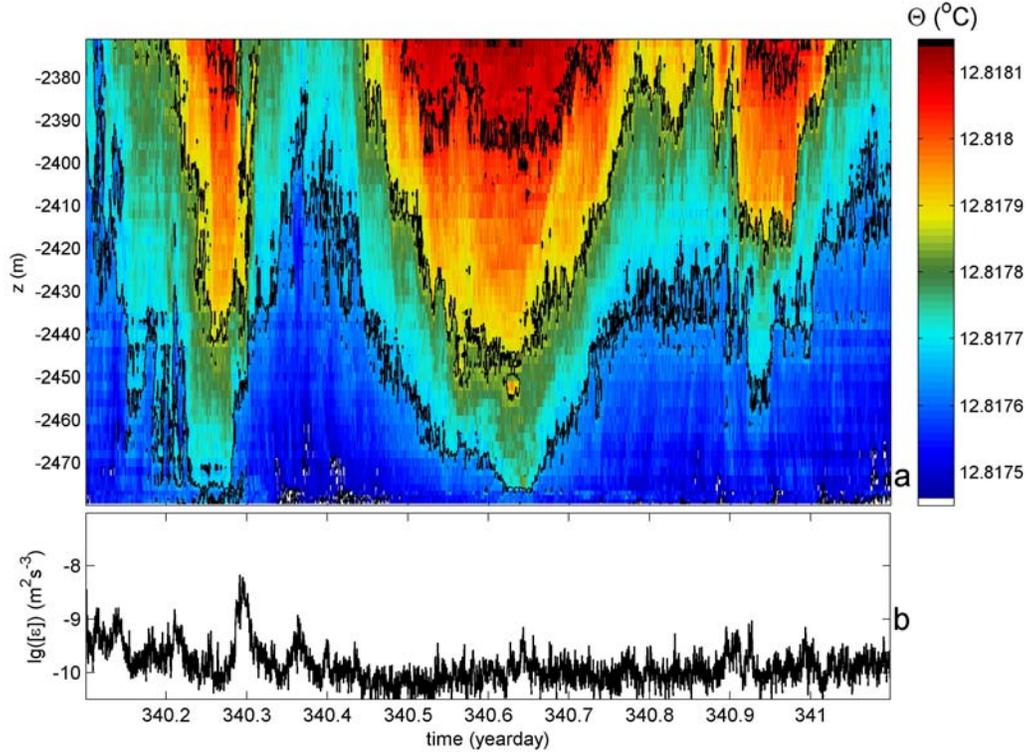

**Figure 5**. Example of 1-day detailed T-sensor corner-line-1 observations of apparent stable stratification and deep internal waves in late-autumn. (a) Time-vertical image of Conservative Temperature. Black contours are drawn every 0.2 mK. The seafloor is at the x-axis, so that the lowest T-sensor is at 0.5 m above it. (b) Time series of logarithm of vertically averaged turbulence dissipation rate calculated from the data in a.

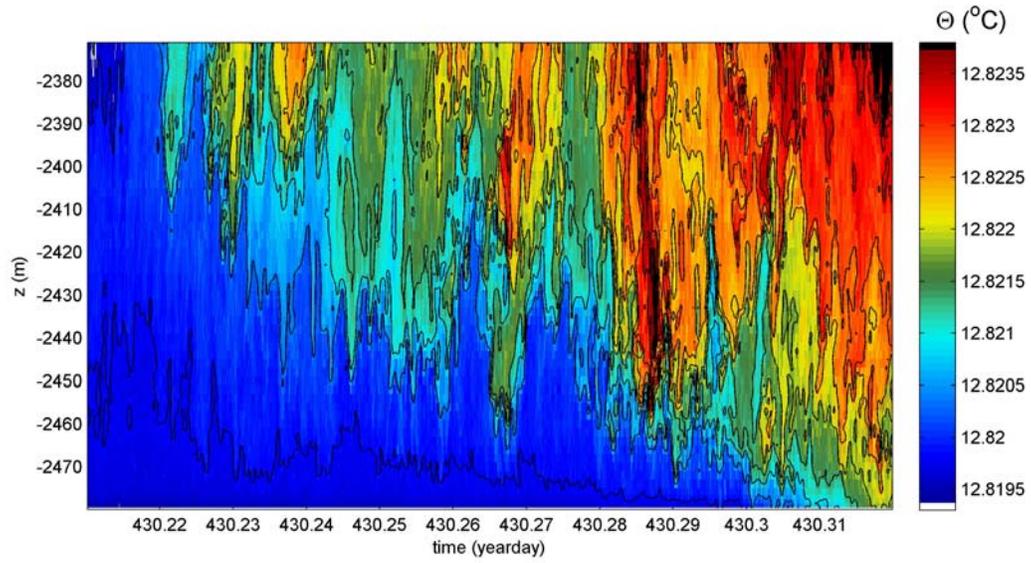

**Figure 6**. As Figure 5a, but for $10^4$ s of observations of apparent stable stratification and convection plumes from above in late-winter. Note the different colour-scale.

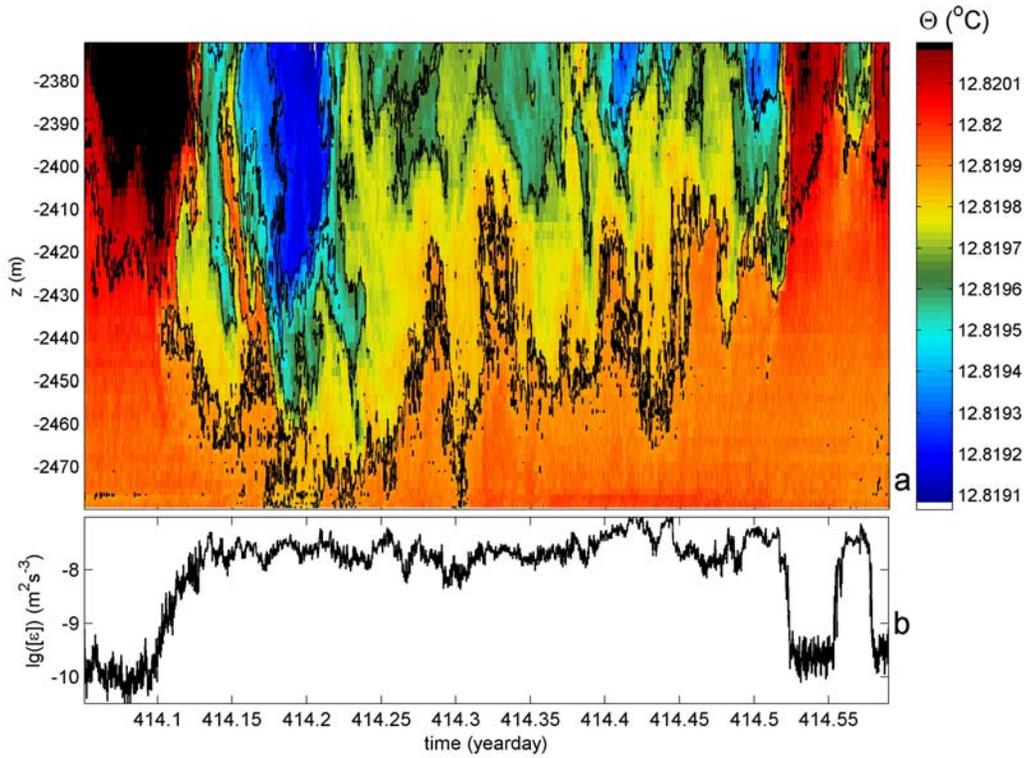

**Figure 7**. As Figure 5, but for 0.54 d of observations including 0.38-d (half inertial period) instability and apparent convection from below, geothermal heating, and relatively warm stratified convection from above on the sides in winter. Note the different colour scale in a.